\documentstyle[seceq,twocolumn,epsf]{jpsj}
\def\H{{\cal H}}
\def\v#1{\mib #1}

\def\ln{\mbox{ln}}
\def\const{\mbox{const}}
\def\JA{J_{\mbox{A}}}
\def\JB{J_{\mbox{B}}}

\def\runtitle{DMRG Study of the $S=1/2$ Quasiperiodic XXZ Antiferromagnets
}

\def\runauthor{Kazuo {\sc Hida}}

\title
{
Density Matrix Renormalization Group Study of the $S=1/2$ Anisotropic Antiferromagnetic Heisenberg Chains with Quasiperiodic Exchange Modulation}

\author
{Kazuo {\sc Hida}
\footnote{E-mail: hida@riron.ged.saitama-u.ac.jp}}

\inst
{
Department of Physics, Faculty of Science,\\ Saitama University, Urawa, Saitama 338-8570
}

\recdate
{
}

\abst
{
The low energy behavior of the $S=1/2$ antiferromagnetic XY-like XXZ chains with precious mean quasiperiodic exchange modulation is studied by the density matrix renormalization group method. It is found that the energy gap of the chain with length $N$ scales as $\exp (-cN^{\omega})$ with nonuniversal exponent $\omega$ if the Ising component of the exhange coupling is antiferromagnetic. This behavior is expected to be the characteristic feature of the quantum spin chains with relevant aperiodicity. This is in contrast to the XY chain for which the precious mean exchange modulation is marginal and the gap scales as $N^{-z}$. On the contrary, it is also verified that the energy gap scales as $N^{-1}$ if the Ising component of the exhange coupling is ferromagnetic. Our results are not only consistent with the recent bosonization analysis of Vidal, Mouhanna and Giamarchi\cite{vidal} but also clarify the nature of the strong coupling regime which is inaccesssible by the bosonization approach.}

\kword
{
quasiperiodicity, XXZ chain, precious mean chain, density matrix renormalization group
}
\begin{document}
\sloppy
\maketitle
\section{Introduction}
The low energy properties of the quantum spin systems with modulated spatial structure have been attracting broad interest in recent studies of quantum many body problem. Although the periodic and the random chains are studied in detail, the quasiperiodic chains, which has an intermediate character between the regular and random chains, are less studied except for the XY-case which is equivalent to the spinless free fermion chains\cite{kkt1,oprss,ko1,kst1,hk1,jh1}. In the fermionic language, the Ising component of the exchange coupling corresponds to the fermion-fermion interaction leading to the strong correlation effect which is the most important subject of the recent condensed matter physics. 

Although the XY chain can be mapped onto the free fermion chain, the problem is not trivial on the quasiperiodic lattice. For the Fibonacci lattice, Kohmoto and coworkers\cite{kst1,hk1}  clarified the Cantor-set structure of the single particle spectrum and the wave function by means of the renormalization group (RG) method. Especially the dynamical exponents are found analytically at the band center and the band edge. Recently, this approach has been extended to include other types of quasiperiodic lattices and the anisotropy between the $x$ and $y$ component of exchange couplings\cite{jh1}. It should be noted that the criticality of the Fibonacci XY chain stems from the marginal nature of the Fibonacci and other precious mean aperiodicity. For the relevant aperiodicity, more singular behavior with divergent dynamical exponent is realized even for the XY chain\cite{jh1}. 

As another example of spin chains which can be mapped onto the free fermion systems, the aperiodic transverse field Ising chains have been studied extensively\cite{jh2,ig1,ig2}. It is known that the aperiodic modulation of the exchange and/or transverse field can be relevant, marginal or irrelevant depending on the substitution rules which generate the aperiodic sequence. In numerical works, such scaling properties are reflected on the energy gap distribution among the subsequences of the infinite aperiodic chains\cite{ig1,ig2}. 

Although these works have revealed a beautiful mathematical structure of quasiperiodic chains, almost no attempts to include the interaction effect are carried out so far except for the mean field approach\cite{hh1,hk1} and recent bosonization approach\cite{vidal}. In the present work, we employ the density matrix renormalization group (DMRG) method\cite{wh1,kh1} to take full account of the correlation effect in the $S=1/2$ precious mean antiferromagnetic XXZ chains which include the Fibonacci chains.

This paper is organized as follows. In the next section, the model Hamiltonian is presented.  The results of the analytical approaches such as the exact renormaliztion group method for the XY chain and the bosonization argument by Vidal {\it et al.}\cite{vidal} for the XXZ chain are breifly reviewed in section 3. The possible candidates of the scaling properties of the energy gap distribution is presented in section 4 based on the previouly known results for the qusaiperiodic XY chains and random chains. The numerical results are presented and analyzed in section 5. The last section is devoted to summary and discussion. Some of the present results are already reported in ref. \citen{khf}.
\section{Model Hamiltonian}
We consider the quasiperiodic XXZ model given by the Hamiltonian,
\begin{equation}
\label{eq:ham}
\H = \sum_{i=1}^{N-1} 2J_{\alpha_i}\left[ S^x_{i}S^x_{i+1}+S^y_{i}S^y_{i+1} + \Delta S^z_{i}S^z_{i+1} \right],\ \ \ (J_{\alpha_i} > 0),
\end{equation}
where $\v{S}_{i}$'s are the spin 1/2 operators and the open boundary condition is assumed.  The exchange couplings $J_{\alpha_i}$'s ($=\JA$ or $\JB$) follow the precious mean sequence generated by the substitution rule,
\begin{equation}
\label{subst}
A \rightarrow A^k B, \ B \rightarrow A.
\end{equation}
The cases $k=1$ and $k=2$ correspond to the Fibonacci (golden mean) and silver mean chains, respectively. In the following, we take $\JA=1$ to fix the energy unit.

 The cases $\Delta =1$ and $0$ correspond to the antiferromagnetic Heisenberg and XY chains, respectively. For $\Delta \leq -1$, it is obvious that the ground state of this model is the fully polarized ferromagnetic state because each bond prefers the ferromagnetic configuration. On the other hand, for $\Delta >1$, the N\'eel type antiferromagnetic order is expected, beacuse each bond has antiferromagnetic Ising type anisotropy.  In the following, we therefore concentrate ourselves on the XY-like regime ($1 \geq \Delta > -1$).

\section{Analytical Approach}

\subsection{Exact Renormalization Group Approach for the XY Chains}
For the Fibonacci XY chains, Kohmoto and coworkers\cite{ko1,kst1} formulated the exact renormalization group for the transfer matrices and calculated the scaling properties of the energy spectrum. Recently, this approach has been extended to include the case of general aperiodic chains by Hermisson\cite{jh1}. It was found that the energy scale $\Delta E$ of the finite size aperiodic XY chain scales with the system size $N$ as 
\begin{equation}
\Delta E \sim N^{-1},
\end{equation}
\begin{equation}
\Delta E \sim N^{-z},
\end{equation}
or
\begin{equation}
\Delta E \sim \exp(-cN^{\omega}),
\end{equation}
according as whether the aperiodic modulation of the exchange coupling is irrelevant, marginal or relevant under the renormalization group transformation. The constants $z$ and $\omega$ are non-universal exponents.

Among them, the precious mean  XY chain belongs to the marginal case and the exponent $z$ is given by,\cite{ko1,kst1,hk1,jh1}
\begin{equation}
\label{dexp}
z=\frac{\ln (f^2 + \sqrt{1+f^2})}{3\ln \phi},
\end{equation}
with
\begin{displaymath}
f = \frac{1}{\sqrt{2}}\left(\frac{\JA}{\JB}+\frac{\JB}{\JA}\right),\ \phi=\frac{1+\sqrt{5}}{2}.
\end{displaymath}
for the Fibonacci chains.

\subsection{Bosonization and Renormalization Group Approach for the XXZ Chains}

The interacting spinless fermion chain with Fibonacci potential has been studied by Vidal et al.\cite{vidal} by means of the bosonization and RG technique. This model can be mapped onto the XXZ chain in the Fibonacci magnetic field by the Jordan-Wigner transformation.  After bosonization, this model can be described by the boson Hamiltonian

\begin{equation} \label{Hbos}
H=H_0+H_W^h,
\end{equation}
\begin{equation} \label{H0bos}
H_0={1\over 2\pi}\int dx \left[(u K)(\pi \Pi)^2+\left({u\over K}\right)
(\partial_x \phi)^2\right],
\end{equation}
\begin{equation}
H_W^h = \frac1{2\pi \alpha} \int dx \,W(x)  \cos\left[2k_Fx+\sqrt{2}\phi(x)\right],
\label{hquasi}
\end{equation}
where $\phi$ is the boson field defined in the interval $[0, \sqrt{2}\pi]$, $\Pi$, the momentum field conjugate to $\phi$, $\alpha$, the ultraviolet cut-off, $u$, the spin wave velocity, $k_F$, the fermi wave number of the spinless fermions and $K$, the Luttinger liquid parameter. The function $W(x)$ represents the spatially varying magnetic field.  For the Fibonacci type modulation, the function $W(x)$ is defined via its Fourier components given in ref. \citen{vidal}. Using the standard bosonization scheme, the spatial modulation of the exchange coupling is similarly expressed as,
\begin{equation}
H_W^J = \frac1{2\pi \alpha} \int dx \, W(x)  \sin\left[2k_Fx+\sqrt{2}\phi(x)\right],
\end{equation}
which coincides with eq. (\ref{hquasi}) by the shift of the origin of $\phi$. Therefore the conclusion obtained by Vidal {\it et al.} also holds for the case of Fibonacci type exchange modulation given by the Hamiltonian (\ref{eq:ham}).

The Luttinger liquid parameter $K$ can be calculated by the Bethe's hypothesis\cite{bik,lp} for the uniform chain as
\begin{equation}
K=\frac{\pi}{\pi-\mbox{arccos}\Delta}.
\label{bethe}
\end{equation}
The case $K=1$ corresponds to the $SU(2)$ invariant isotropic Heisenberg chain ($\Delta = 1$) and $K=2$ to the XY chain ($\Delta=0$). (Note that our definition of $K$ differs from that of ref. \citen{vidal} by a factor of 2.)

Vidal {\it et al.}\cite{vidal} derived the RG equation within the weak coupling approximation. Based on the numerical solution of the RG equation, they obtained the following results.  For $K > K_c \simeq 2$, the Fibonacci modulation is irrelevant and the ground state is the usual Luttinger liquid. On the other hand, the Fibonacci modulation is relevant for $K < K_c$ and the ground state is renormalized to the strong coupling regime. In this case, the weak coupling theory cannnot predict the ground states properties.  For $K=K_c$, the Fibonacci modulation becomes marginal and the ground state is critical with nonuniversal dynamical exponent $z$. This is consistent with the well-known case of the free spinless fermions in the Fibonacci potential if $K_c$ exactly equals 2, because $K=2$ corresponds to $\Delta=0$ by eq. (\ref{bethe}). We therefore assume $K_c=2$ exactly in what follows. Thus the XY-like XXZ chain with antiferromagnetic Ising coupling ($1 \leq K < 2; 0 < \Delta \leq 1$) is renormalized to the strong coupling regime while that with ferromagnetic Ising coupling ($2 < K; -1 < \Delta < 0$) is renormalized to the normal Luttinger liquid. In what follows, we mainly aim to clarify the nature of the ground state in the strong coupling regime $0 < \Delta \leq 1$, which is inaccessible by the weak coupling approximation, with the help of numerical DMRG method. At the same time, we also intend to verify the prediction of the RG calculation\cite{vidal} by numerical calculation.

\begin{figure}
\epsfxsize=70mm 
\centerline{\epsfbox{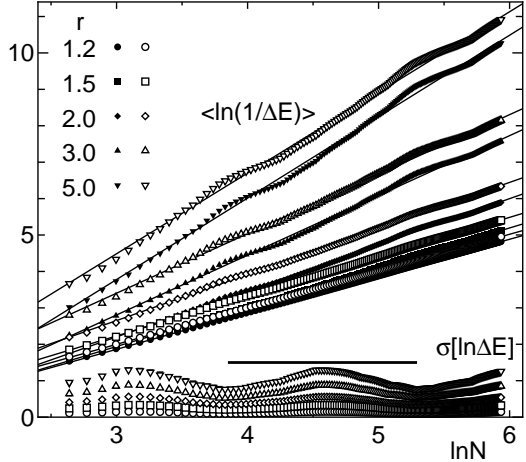}}
\caption{The $N$-dependence of $<\ln(1/\Delta E)>$ and $\sigma[\ln\Delta E]$  for the Fibonacci XY chain by the exact diagonalization method. The length of the horizontal bar is $3\ln((1+\sqrt{5})/2)$.In this and following figures \ref{fig3}, \ref{figfit1}, \ref{fig5}, \ref{fig32xsz}, \ref{figfit2}, \ref{fig6}, \ref{fibxfitx}, \ref{fig12} and \ref{fibxf}, $r \equiv \mbox{Max} \{ \JB/\JA, \JA/\JB \}$  and the filled (open) symbols represent the case $\JB > \JA (\JB < \JA)$. The filled symbols for $\sigma$ almost overlap with open symbols.}
\label{fig1}
\end{figure}

\begin{figure}
\epsfxsize=70mm 
\centerline{\epsfbox{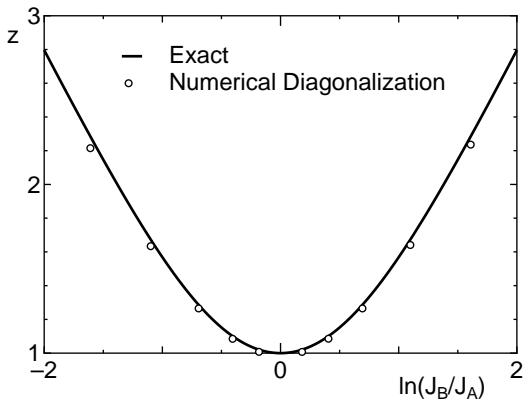}}
\caption{The dynamical exponent $z$ for the XY chain. The filled circle represent the present numerical calculation and solid line is the analytical results by Kohmoto {\it et al}.\cite{ko1,kst1}}
\label{fig2}
\end{figure}

\section{Scaling Properties of Energy Gap Distribution}

In order to clarify the bulk properties of the infinite precious mean chains from the finite size calculation, it is not sufficient to investigate the properties of a single series of precious mean chains generated by the substitution rule (\ref{subst}). The precious mean sequence of length $n$ ($ABA$ for $n=3$ or $ABAAB$ for $n=5$ in the Fibonacci case) is always the $n$-membered  subsequence at the beginning of the infinite precious mean sequence ($ABAABABAABAAB....$ in the Fibonacci case).  The properties of the finite length precious mean chains therefore correspond to the  surface properties of the infinite precious mean chains rather than its bulk properties. Hence, it is necessary to consider all possible $n$-membered subsequences of the infinite precious mean chain and investigate the distribution of their physical properties, such as energy gaps, to reveal the {\it bulk} properties of the infinite precious mean chains. This method has been successfully applied to the quasiperiodic transverse field Ising chains\cite{ig2}. It should be noted that the number of the $n$-membered subsequence is equal to $n+1$\cite{penrose}.

The scaling property of the energy gap distribution is determined by the system size dependence of the characteristic energy scale. In the XY case, the precious mean aperiodicity in the exchange coupling is marginal and the energy gap $\Delta E$ scales with the system size as $\Delta E \sim N^{-z}$ where $z$ is the dynamical exponent\cite{jh1} as discussed in the preceding section. Therefore the gap distribution function scales as 
\begin{equation}
\label{cr}
P(\Delta E)d\Delta E = N^z f( N^z \Delta E) d\Delta E.  
\end{equation}
Consequently, the average and flucutation of $\ln(1/\Delta E)$ scale as,
\begin{eqnarray}
<\ln(1/\Delta E)> &\simeq& C_1-z \ln N, \\ 
\sigma[\ln(\Delta E)] &\equiv& \sqrt{<(\ln(\Delta E)- <\ln(\Delta E)>)^2>}, \nonumber \\
 &\simeq& \sqrt{C_2-C_1^2} = \const., 
\end{eqnarray}
where
\begin{displaymath}
C_n = \int_{-\infty}^{\infty}t^n g(t)dt, \ g(t)=f(\mbox{e}^t). 
\end{displaymath}
This type of behavior is also observed in the Griffith phase of the random quantum spin chains such as $S=1/2$ random dimerized Heisenberg chains\cite{hy1,kh2} or $S=1$ random Heisenberg chains\cite{bsc1,hy2,mon1,yn1,yn2,tkt1,khh}
 
On the other hand, for the XY chain with relevant exchange aperiodicity, the gap scales as $\ln (1/\Delta E) \sim N^{\omega}$\cite{jh1} and the gap distribution function scales as, 
\begin{equation}
\label{rs}
P(\ln\Delta E)d\ln\Delta E = N^{-\omega}f(N^{-\omega}\ln\Delta E) d\ln\Delta E,  
\end{equation}
which gives
\begin{eqnarray}
\label{rel}
<\ln(1/\Delta E)> &\simeq& D_1 N^{\omega}, \\
\sigma[\ln(\Delta E)] &\simeq& \sqrt{D_2-D_1^2} N^{\omega},  
\end{eqnarray}
where
\begin{displaymath}
D_n = \int_{-\infty}^{\infty}x^n f(x)dx. 
\end{displaymath}
It should be remarked that $\sigma$ tends to a constant value for the marginal aperiodicity while it grows with the same exponent as $<\ln(1/\Delta E)>$ for the relevant aperiodicity. This type of behavior with $\omega=1/2$ is observed also in the random singlet phase of the $S=1/2$ random antiferromagnetic XXZ chain.\cite{kh1,ds1,mckenzie}

\section{Numerical Results}
\subsection{XY chains ($\Delta=0$)}

In the XY chain, the energy spectrum can be calculated by numerical diagonalization of $N \times N$ matrices. The average $<\ln(1/\Delta E)>$ and fluctuation $\sigma[\ln\Delta E]$ of the logarithm of the energy gap for all possible $(N-1)$-membered subsequence is calculated for $14 \leq N \leq 378$ and various values of $\JB/\JA$ between 1/5 and 5. These are plotted against $\ln N$ in Fig. \ref{fig1}. The linearity of  $<\ln(1/\Delta E)>$ to  $\ln N$ is fairly good and the flucutation $\sigma$ also tends to a constant value except for the oscillation with period $3 \ln\{(1+\sqrt{5})/2\}$ which is inherent to the Fibonacci chain spectrum\cite{luck,jh1}. The facter 3 comes from the fact that the single step of the RG transformation for the precious mean chains with odd $k$ corresponds to three inflation steps\cite{jh1}. We further calculated the dynamical exponent $z$ by fitting $\ln(1/\Delta E)$-$\ln N$ curve by a straight line. The obtained values of $z$ are shown in Fig. \ref{fig2} by open circles. The solid line is the exact expression (\ref{dexp}). Again we find a good agreement. This result also confirms that the energy gap distribution of the finite length subsequences of the Fibonacci chain correctly reflects scaling properties of the ground state.
  
\begin{figure}
\epsfxsize=70mm 
\centerline{\epsfbox{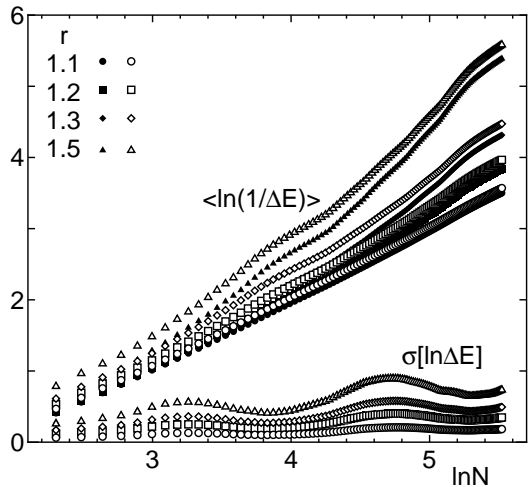}}
\caption{The $N$-dependence of $<\ln(1/\Delta E)>$ and $\sigma[\ln\Delta E]$ for the Fibonacci Heisenberg chain by the DMRG method plotted against $\ln N$.  }
\label{fig3}
\end{figure}

\begin{figure}
\epsfxsize=70mm 
\centerline{\epsfbox{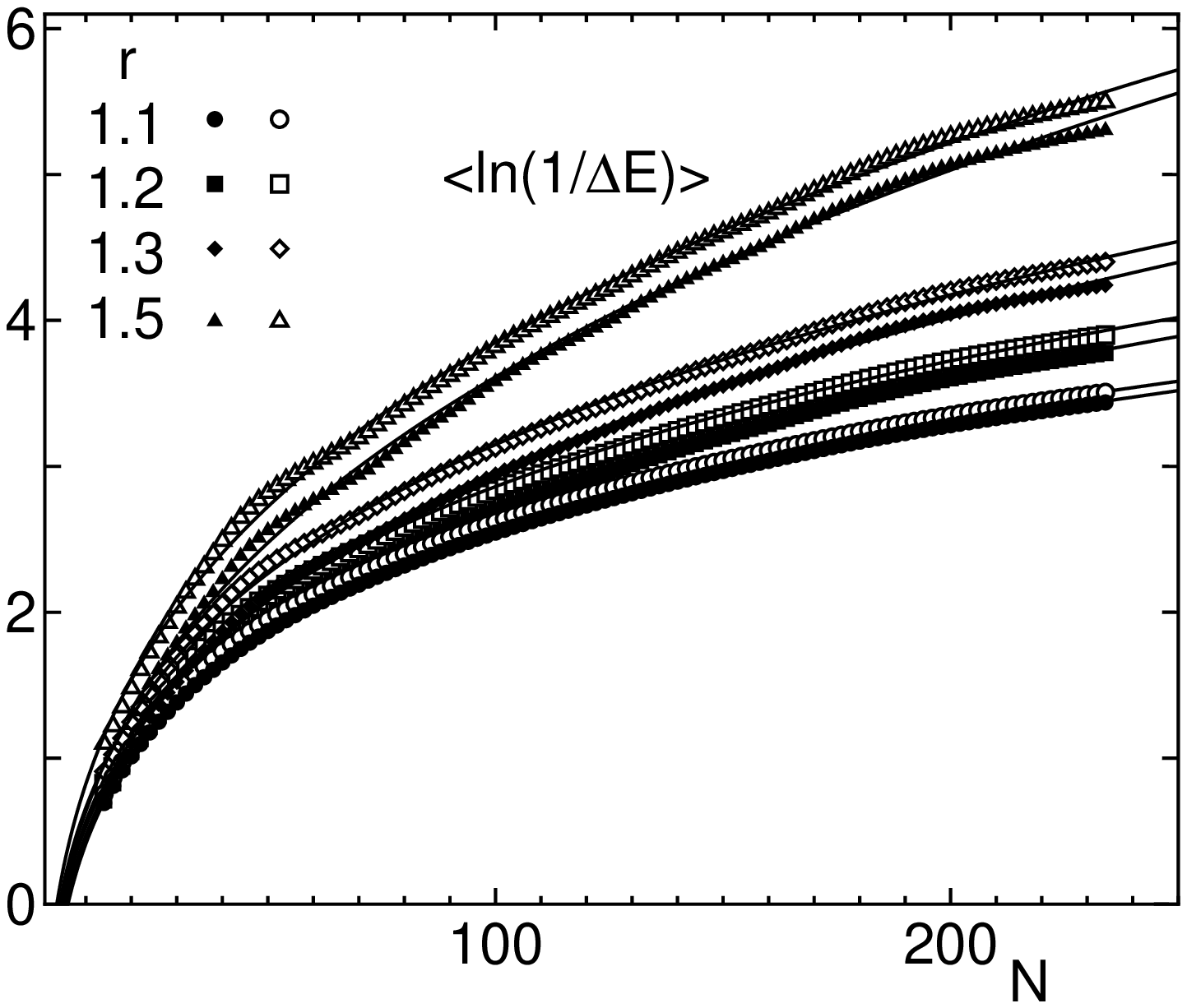}}
\caption{The $N$-dependence of $<\ln(1/\Delta E)>$ for the Fibonacci Heisenberg chain by the DMRG method plotted against $N$. The solid surve is the fitting by the formula $<\ln(1/\Delta E)>=D_1N^{\omega} + \const.$}
\label{figfit1}
\end{figure}

\begin{figure}
\epsfxsize=70mm 
\centerline{\epsfbox{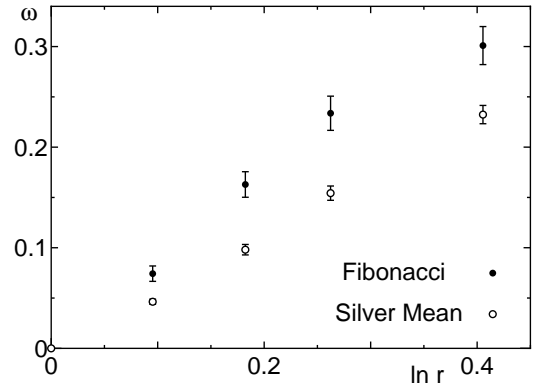}}
\caption{The exponent $\omega$ for the Fibonacci  and silver mean Heisenberg chains.}
\label{fig4}
\end{figure}
\begin{figure}
\epsfxsize=70mm 
\centerline{\epsfbox{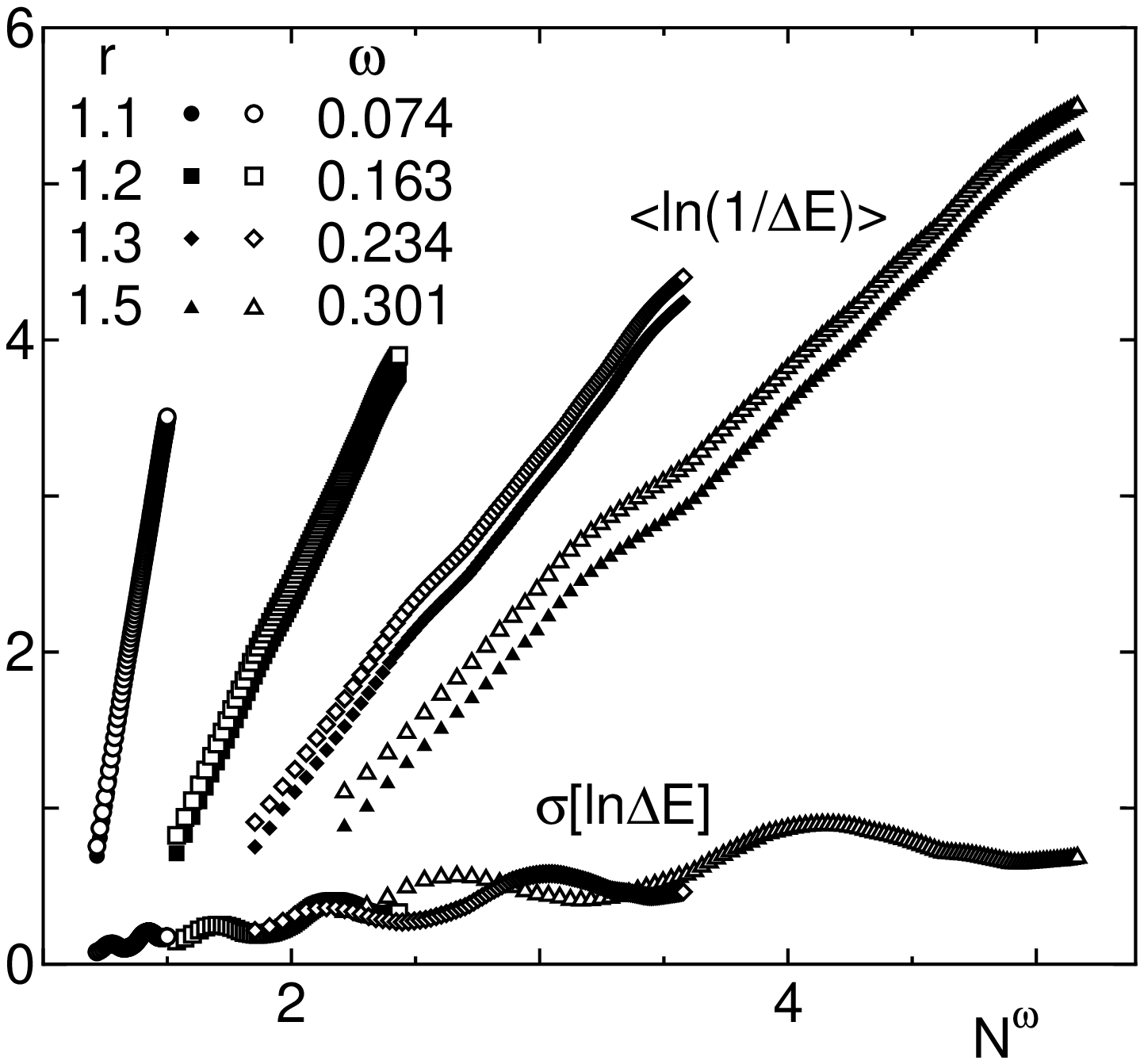}}
\caption{The $N$-dependence of $<\ln(1/\Delta E)>$ and $\sigma[\ln\Delta E]$ for the Fibonacci Heisenberg chain by the DMRG method plotted against $N^{\omega}$.}
\label{fig5}
\end{figure}
\begin{figure}
\epsfxsize=70mm 
\centerline{\epsfbox{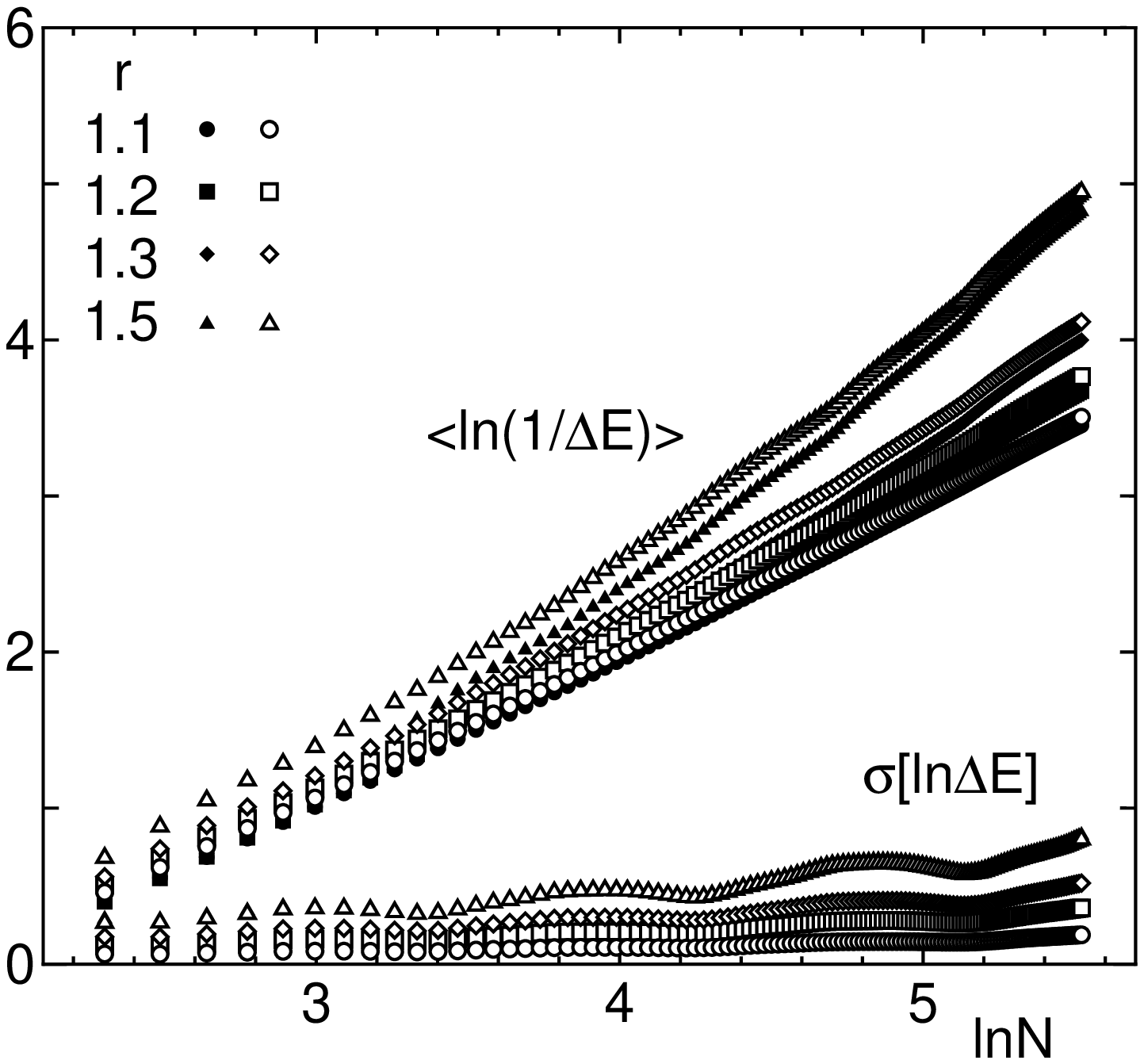}}
\caption{The $N$-dependence of $<\ln(1/\Delta E)>$ and $\sigma[\ln\Delta E]$ for the silver mean Heisenberg chain by the DMRG method plotted against $\ln N$.  }
\label{fig32xsz}
\end{figure}
\begin{figure}
\epsfxsize=70mm 
\centerline{\epsfbox{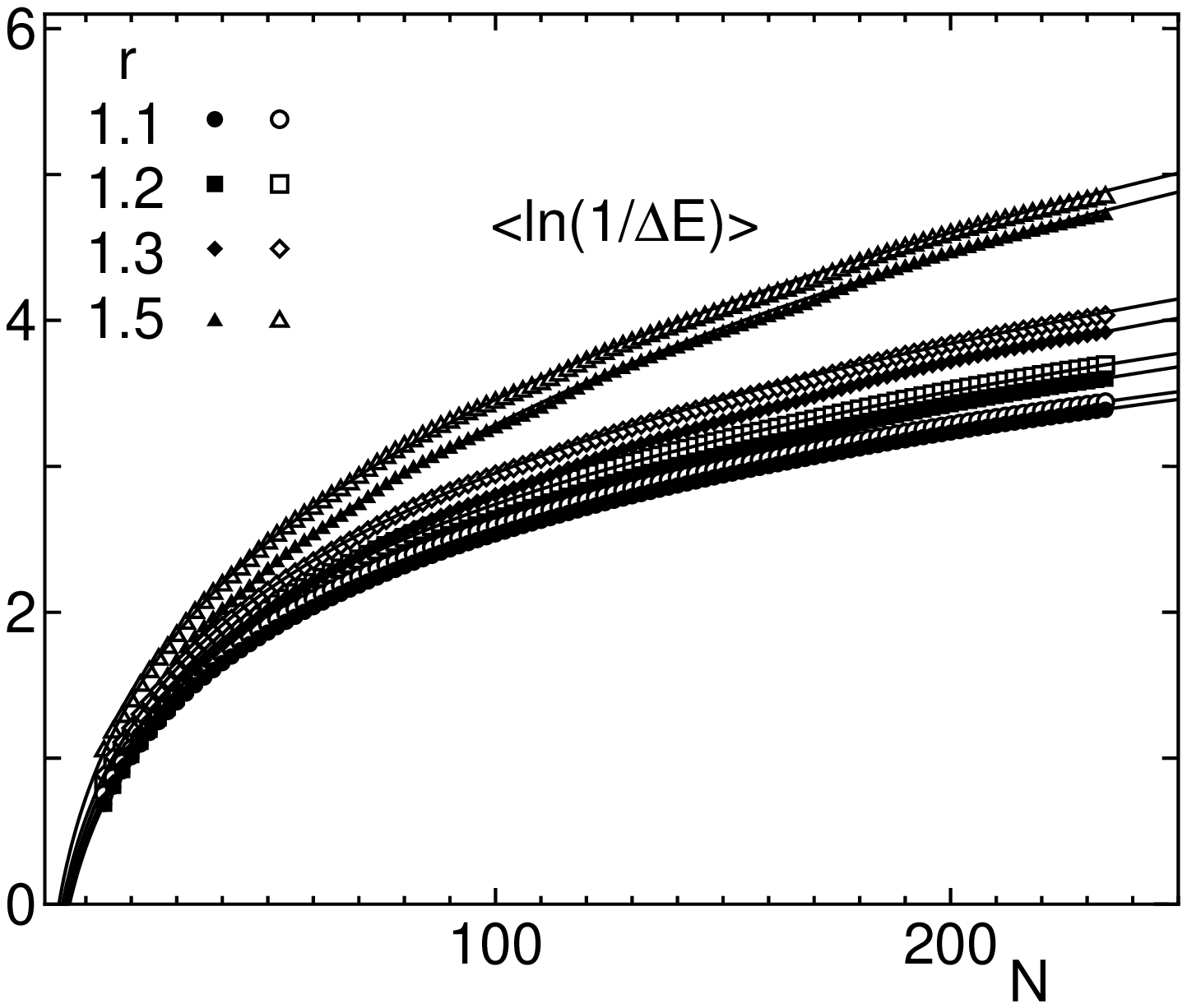}}
\caption{The $N$-dependence of $<\ln(1/\Delta E)>$ for the silver mean Heisenberg chain by the DMRG method plotted against $N$. The solid surve is the fitting by the formula $<\ln(1/\Delta E)>=D_1N^{\omega} + \const$.}
\label{figfit2}
\end{figure}
\begin{figure}
\epsfxsize=70mm 
\centerline{\epsfbox{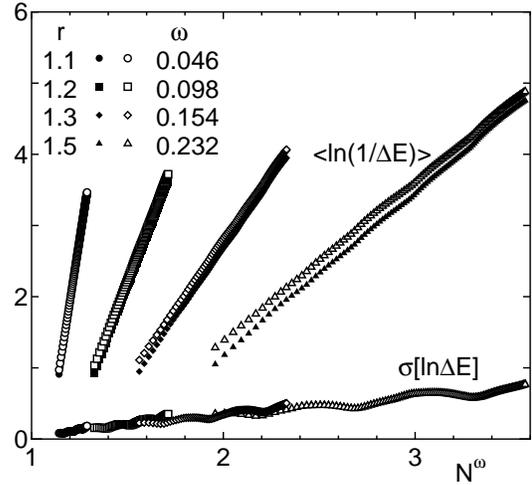}}
\caption{The $N$-dependence of $<\ln(1/\Delta E)>$ and $\sigma[\ln\Delta E]$ for the silver mean Heisenberg chain by the DMRG method plotted against $N^{\omega}$.}
\label{fig6}
\end{figure}
\subsection{Isotropic Heisenberg Chains ($\Delta=1$)}
The energy gap distribution for the Fibonacci Heisenberg chain ($\Delta=1$) is calculated by the DMRG method for  $14 \leq N \leq 234$ and various values of $\JB/\JA$ between 2/3 and 3/2 using the algorithm developed for the random chains\cite{kh1}. The number $m$ of the states kept at each step of DMRG iteration is 60 and only infinite size iterations are carried out. We have checked that $m$-dependence is negligible by increasing $m$ up to 80 for $J_B/J_A =3/2$ which is the most dangerous case studied here. For the further check of the accuracy of the DMRG scheme, we have also calculated the energy spectrum of the XY chain with $14 \leq N \leq 234$ using DMRG and found that the results coincide with the exact diagonalization results within the size of the symbols of Fig. \ref{fig1}.

 Figure \ref{fig3} shows the average $<\ln(1/\Delta E)>$ and fluctuation $\sigma[\Delta E]$ plotted against $\ln N$ for the Fibonacci Heisenberg chain.  The curves of  $<\ln(1/\Delta E)>$ show an evident upturn as $N$ increases. Further, the fluctuation $\sigma$ does not tend to a constant value. We have fitted the data for $<\ln(1/\Delta E)>$ by the power law 
\begin{equation}
<\ln(1/\Delta E)>=D_1N^{\omega} + \const. ,
\end{equation}
as shown in Fig. \ref{figfit1}. The constant term is added to (\ref{rel}) to account for the arbitrariness of the energy scale. The exponent $\omega$ turned out to be non-universal depending on the ratio $\JB/\JA$ as $\omega=\omega(\JB/\JA)$. We have further assumed that $\omega(\JB/\JA)=\omega(\JA/\JB)$ because the RG equation of ref. \citen{vidal} is invariant under the exchange $\JA \leftrightarrow \JB$. It should be also noted that the dynamical exponent for the precious mean XY chain is also invariant under the exchange $\JA \leftrightarrow \JB$. In addition, the numerically obtained values of $\omega$ for the silver mean chains also satisfiy the relation  $\omega(\JB/\JA) \simeq \omega(\JA/\JB)$ as explained below. So we assume that this relation also holds for the Fibonacci Heisenberg chains and use the average of the numerically obtained values of $\omega(\JB/\JA)$ and $\omega(\JA/\JB)$ as $\omega$. The values of $\omega$ are depicted against $\ln r$ ($r \equiv \mbox{Max}(\JB/\JA,\JA/\JB)$) in Fig. \ref{fig4} by filled symbols. The error bars in Fig. \ref{fig4} are estimated from the difference between $\omega(\JB/\JA)$ and $\omega(\JA/\JB)$. Using thus obtained values of $\omega$,  $<\ln(1/\Delta E)>$ and $\sigma[\ln\Delta E]$ are plotted against $N^{\omega}$ in Fig. \ref{fig5}.  It is clearly seen that both  $<\ln(1/\Delta E)>$ and $\sigma[\ln\Delta E]$ grow linearly with $N^{\omega}$.

The same analysis is made for the silver mean chain in Fig. \ref{fig32xsz}, Fig. \ref{figfit2} and Fig. \ref{fig6} for $18 \leq N \leq 240$. In this case, the oscillation period in $\ln N$ is $\ln(1+\sqrt{2})$, because the single step of the RG transformation for the precious mean chains with even $k$ corresponds to a single inflation step\cite{jh1}. The numerically obtained values of $\omega$ satisfy $\omega(\JA/\JB)\simeq\omega(\JB/\JA)$ with better accuracy than the Fibonacci case. The values of $\omega$ are plotted against $\ln r$ in Fig. \ref{fig4} by open symbols. 

The analysis in this section clearly indicates that the universality class of the precious mean isotropic Heisenberg chain is different from that of the corresponding XY chain. To further clarify the origin of the difference between these two types of behavior, we next investigate the case of XXZ chain which interpolates between these two limiting cases.

\subsection{XY-like XXZ chains ($-1 < \Delta \leq 1$)}
The same analysis is carried out for the Fibonacci XXZ chain with $-1 < \Delta \leq 1$ for $r = 1.2$.  The energy gap is estimated from the difference between the ground state energies with $S^z_{\rm tot}=0$ and 1. The number of the states kept in each DMRG step was 70. 

\begin{figure}
\epsfxsize=70mm 
\centerline{\epsfbox{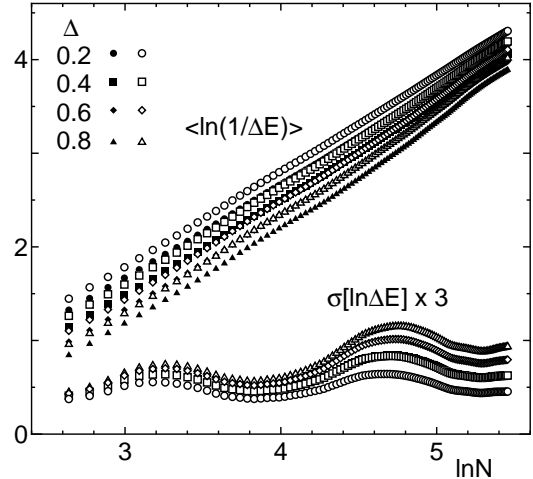}}
\caption{The $N$-dependence of $<\ln(1/\Delta E)>$ and $\sigma[\ln\Delta E]$ for the Fibonacci XXZ chain by the DMRG method plotted against $\ln N$.  }
\label{fig33xsz}
\end{figure}

\begin{figure}
\epsfxsize=70mm 
\centerline{\epsfbox{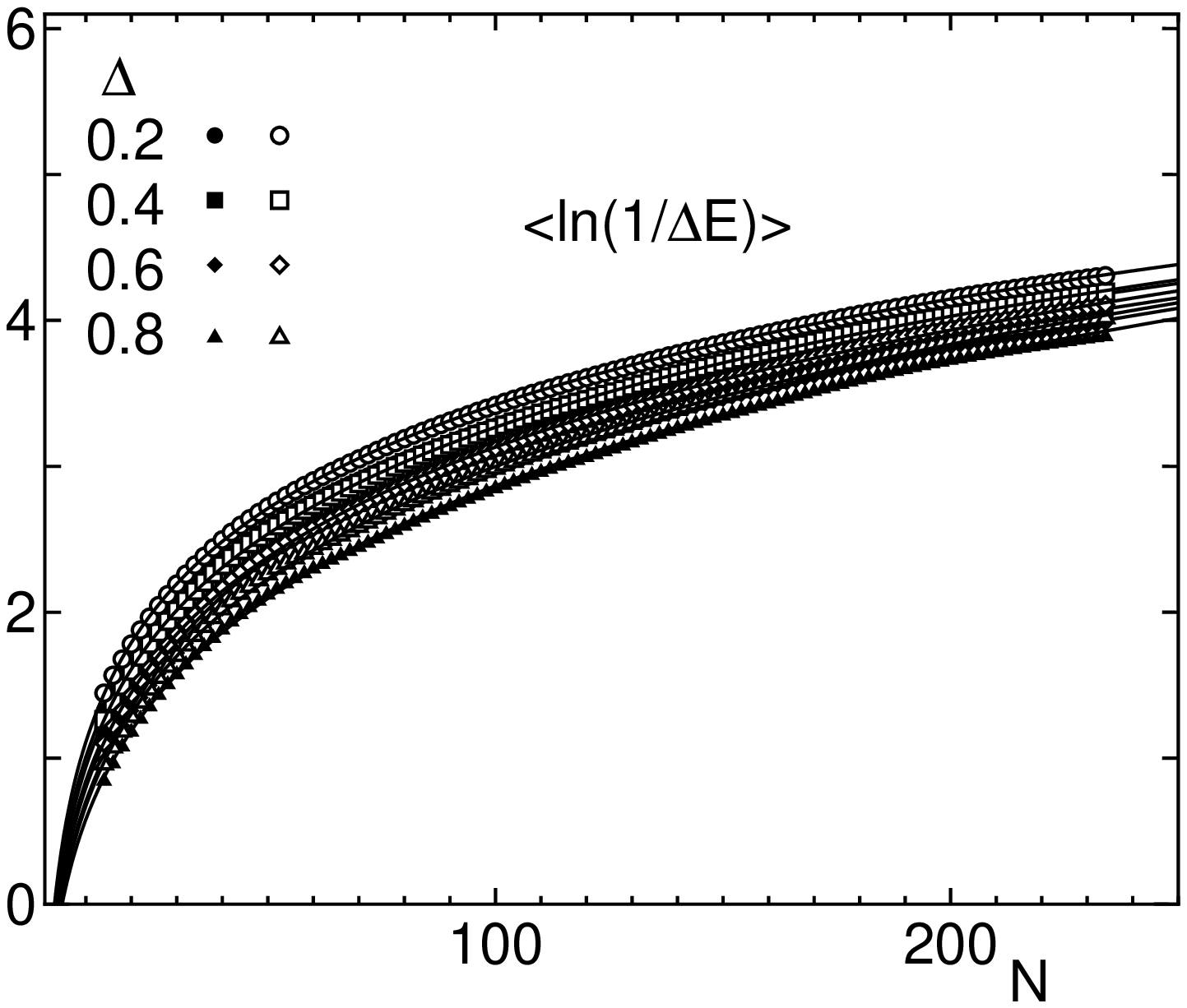}}
\caption{The $N$-dependence of $<\ln(1/\Delta E)>$ for the Fibonacci XXZ chain by the DMRG method plotted against $N$. The solid surve is the fitting by the formula $<\ln(1/\Delta E)>=D_1N^{\omega} + \const$.}
\label{fibxfitx}
\end{figure}

\begin{figure}
\epsfxsize=70mm 
\centerline{\epsfbox{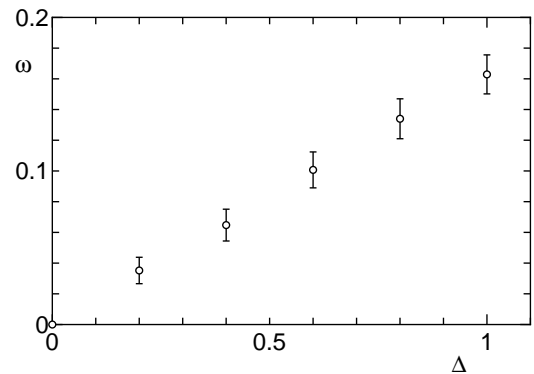}}
\caption{The anisotropy dependence of the exponent $\omega$ for the Fibonacci  XXZ chains with $r=1.2$.}
\label{fig7}
\end{figure}

\begin{figure}
\epsfxsize=70mm 
\centerline{\epsfbox{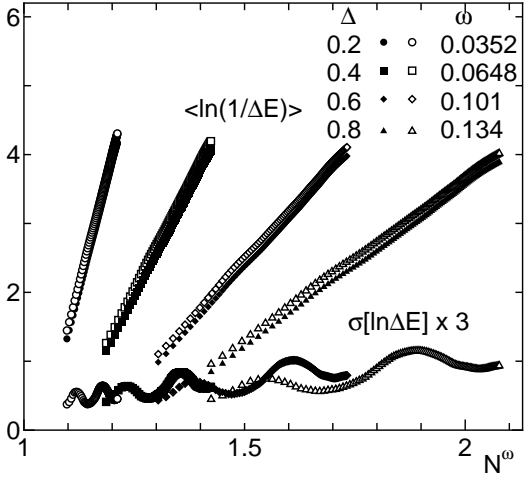}}
\caption{The $N$-dependence of $<\ln(1/\Delta E)>$ and $\sigma[\ln\Delta E]$ for the Fibonacci XXZ chain with $0 \leq \Delta \leq 1$ by the DMRG method plotted against $N^{\omega}$.}
\label{fig12}
\end{figure}

Let us first discuss the antiferromagnetic XXZ regime $0 < \Delta \leq 1$. Figure \ref{fig33xsz} shows the plot of  the average $<\ln(1/\Delta E)>$ and fluctuation $\sigma[\Delta E]$ against $\ln N$. In spite of the apparently linear behavior for small $\Delta$, the upturn becomes evident as $\Delta$ approached unity. Figure \ref{fibxfitx} shows the fitting procedure and  Fig. \ref{fig7}, the $\Delta$ dependence of $\omega$.  Although the precise numerical estimation of $\omega$ for small $\Delta$ is  rather difficult, it is clearly seen that $\omega$ takes finite values for nonzero $\Delta$ and  decreases almost linearly with $\Delta$.  The quantities $<\ln(1/\Delta E)>$ and $\sigma[\ln\Delta E]$ are plotted against $N^{\omega}$ in Fig. \ref{fig12}. This behavior of the energy gap clearly indicates that the gap distribution of the type (\ref{rs}) with finite $\omega$ is generated as soon as the correlation effect is switched on. The precious mean XXZ antiferromagnetic chains with $1 \geq \Delta >0 $ therefore belongs to the same universality class as the precious mean isotropic Heisenberg chain studied in the preceding subsection. The precious mean aperiodicity is therefore relevant in this regime. 

\begin{figure}
\epsfxsize=70mm 
\centerline{\epsfbox{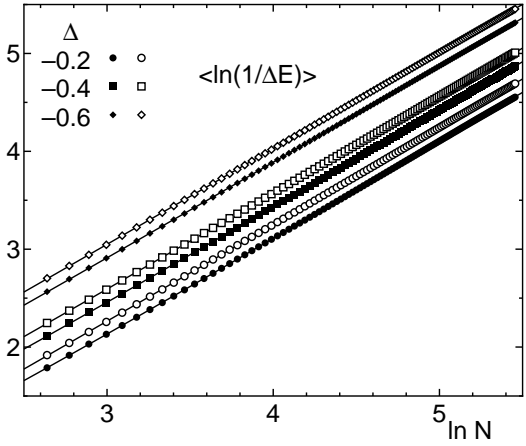}}
\caption{The $N$-dependence of $<\ln(1/\Delta E)>$ for the Fibonacci XXZ chain  with $-1 < \Delta < 0$ by the DMRG method plotted against $\ln N$. }
\label{fibxf}
\end{figure}

On the contrary,  for  $-1 < \Delta < 0$, the data of   $<\ln(1/\Delta E)>$ are well fitted by the straight line $<\ln(1/\Delta E)> \simeq \ln N + \const.$ as shown in Fig. \ref{fibxf} with $\Delta =-0.2, -0.4 $ and $ -0.6$ for $r=1.2$. In all cases, the slope is equal to unity within the accuracy $\pm 0.02$. Therefore the energy gap scales as $N^{-1}$ and the ground state is a normal Luttinger liquid in which the conformal invariance is recovered. This implies that the precious mean aperiodicity is irrelevant in this regime. 

\begin{figure}
\epsfxsize=70mm 
\centerline{\epsfbox{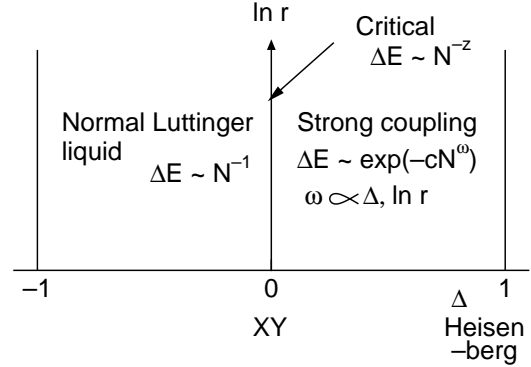}}
\caption{The speculated phase diagram of the precious mean XXZ chain  with $-1 < \Delta \leq 1$. The line $\ln r=0 (r=1)$ corresponds to the uniform XXZ chain whose ground state is the normal Luttinger liquid obviously. }
\label{phase}
\end{figure}

The above behavior is consistent with the conclusion of the RG calculation \cite{vidal} that the XY chain is on the critical point which separates the strong coupling regime $\Delta > 0$ where the precious mean aperiodicity is relevant and weak coupling regime $\Delta < 0$ where the precious mean aperiodicity is irrelevant.\cite{vidal} In addition, our numerical calculation has revealed that the gap distribution in the strong coupling regime is characterized by the scaling form (\ref{rs}) which is similar to that of the XY chain with relevant aperiodic modulation\cite{jh1}. From these observations, the ground state phase diagram of the precious mean XXZ chains are speculated as summarized in Fig. \ref{phase}. It should be remarked in the random exhange case, $S=1/2$ XXZ chains belong to the strong coupling regime for $-1/2 < \Delta \leq 1$\cite{vidal,gia,doty} including the antiferromagnetic Heisenberg and XY points. For these models, the ground state is known to be the random singlet state which is also characterized by the gap distribution of the type (\ref{rs}) with $\omega=1/2$.\cite{kh1,ds1,mckenzie} Presumably, the gap distribution of  type (\ref{rs}) is the generic nature of the aperiodic exchange spin chains in the strong coupling regime. 


\begin{fulltable}
\caption{Energy scale of $S=1/2$ aperiodic spin chains.}
\label{tab1}
\begin{fulltabular}{@{\hspace{\tabcolsep}\extracolsep{\fill}}ccccc} \hline
 & Random XXZ  & Random Dimerized XXZ & Precious Mean XY &Precious Mean XXZ \\ 
 &($-1/2 < \Delta \leq 1$) & ($-1/2 < \Delta \leq 1$) &  & ($0 < \Delta \leq 1$) \\ \hline
Energy Scale& $\exp (-cN^{1/2})$ & $N^{-z}$& $N^{-z}$ & $\exp (-cN^{\omega})$ \\
&  & $z \propto 1/\delta$ & $z$ : (\ref{dexp}) for Fibonacci & $\omega \propto \Delta, \ln r$ \\ \hline
\end{fulltabular}
\end{fulltable}

\section{Summary and Discussion}

The DMRG calculation is carried out for the $S=1/2$ precious mean XY-like XXZ chains. The energy gap distribution of the chains corresponding to all possible $(N-1)$-membered subsequences of the infinite precious mean chains is calculated to clarify the bulk properties of the latter. It is found that the logarithm of the energy gap of the finite size precious mean XXZ chains with $0 < \Delta \leq 1$ scales with a nonuniversal power of the system size as eq. (\ref{rs}).  The exponent $\omega$ is found to increase almost linearly with $\Delta$ and $\mid \ln(\JB/\JA) \mid$. This is distinct from the critical gap distribution described by eq. (\ref{cr}) in the XY case. Our results are not only consistent with the RG calculation by Vidal {\it et al.}\cite{vidal} but also clarify the characteristic features of the low energy spectrum in the strong coupling regime which was unreachable by the weak coupling renormaliation group theory\cite{vidal}. Taking the results for other kinds of aperiodic spin chains\cite{jh1,kh1,ds1,mckenzie,doty} into account, we expect that the energy gap distribution of the type  (\ref{rs}) is characteristic to the aperiodic spin chains in the strong coupling regime where the aperiodicity is relevant. It is remarkable that the well-known critical spectrum of the precious mean XY chain is extremely fragile against the correlation effects. We have also verified that the precious mean aperiodicity is irrelevant for $-1 < \Delta <0$.

On the other hand, in the studies of the random quantum spin chains, it is known that the effect of dimerization drives the random singlet phase in the random XXZ antiferromagnet, in which the randomness is relevant and has the gap distribution (\ref{rs}) with $\omega=1/2$, to the random dimer phase which has the critical gap distribution given by (\ref{cr})\cite{hy1,kh2,mckenzie} where $1/z$ is proportional to the strength of dimerization $\delta$. This implies that  the effect of aperiodic exchange modulation is suppressed by the dimerization while it is enhanced by the antiferromagnetic Ising coupling which induces the correlation effects. This situation is summarized in Table \ref{tab1}.  Thus the competition among the correlation effects, the periodic and aperiodic spatial modulations would produce wide variety of exotic ground states of the aperiodic spin chains which remain to be explored.

In this paper, we concentrated on the case without magnetic field. The effect of the magnetic field is interesting from two different points of view. First, in the presence of the uniform magnetic field, the multifractal Cantor-set structure of the single particle excitation spectrum of the free spinless fermion chain\cite{kst1,luck} manifests itself as the devil's staircase structure of the magnetization curve of the XY chain in the spin language. This can be regarded as the magnetization plateau problem\cite{khp,oya} in the quasiperiodic spatial structure. It is worth investigating if such structure survives in the XXZ case which has stronger quantum fluctuation than the XY case.

Another interesting problem is the effect of the precious mean modulation of magnetic field. For the random field XY chain, the ground stat is known to be localized while it is the random singlet state with divergent spin correlation length for the random exchange XY chain although the randomness is relevant in both cases\cite{doty,mckenzie}. This difference comes from the perfect spin inversion symmetry of the random exchange problem\cite{doty}. From this point of view, the precious mean modulation of the magnetic field would have effects different from that of the exchange coupling in the XXZ chain. This problem is also left for future studies.

The author thanks  M. Kohmoto, T. Ohtsuki and H. Rieger for valuable discussion and comments. The numerical calculations have been performed using the FACOM VPP500 at the Supercomputer Center, Institute for Solid State Physics, University of Tokyo and  the HITAC S820/80 at the Information Processing Center of Saitama University.  This work is supported by the Grant-in-Aid for Scientific Research from the Ministry of Education, Science, Sports and Culture.

\end{document}